\documentclass[12pt]{article}
\usepackage{graphicx}

\begin{document}

\begin{center}

\baselineskip 40pt

\vskip 2cm

{\Large {\bf Quantum critical point of spin-boson model and infrared
catastrophe in bosonic bath}}

\vskip 1cm

{\large {\rm Hang Zheng and Zhiguo L\"{u}}}

Key Laboratory of Artificial Structures and Quantum Control
(Ministry of Education), Department of Physics, Shanghai Jiao Tong
University, Shanghai 200240, China \\

{\bf Abstract}

\end{center}

\baselineskip 20pt

An analytic ground state is proposed for the unbiased spin-boson
Hamiltonian, which is non-Gaussian and beyond the Silbey-Harris
ground state with lower ground state energy. The infrared
catastrophe in Ohmic and sub-Ohmic bosonic bath plays an important
role in determining the degeneracy of the ground state. We show that
the infrared divergence associated with the displacement of the
nonadiabatic modes in bath may be removed from the proposed ground
state for the coupling $\alpha<\alpha_c$. Then $\alpha_c$ is the
quantum critical point of a transition from non-degenerate to
degenerate ground state and our calculated $\alpha_c$ agrees with
previous numerical results.

\vskip 1cm


\pagebreak

\baselineskip 20pt

\section{Introduction}

Quantum impurity systems with competing interactions constitute a
field of wide interest in the quantum physics. In recent years, the
quantum two-level system coupled to dissipative bosonic environment
(spin-boson model, SBM) attracts much attention in this field
because it may be one of the simplest but nontrivial quantum
impurity system for studying the physics of competing interactions.
The Hamiltonian of SBM reads (we set $\hbar =1$)
\begin{equation}
H=-{\frac{1}{2}}\Delta \sigma _{x}+\sum_{k}\omega _{k}b_{k}^{\dag }b_{k}+{\frac{1}{2}}%
\sum_{k}g_{k}(b_{k}^{\dag }+b_{k})\sigma _{z},
\end{equation}
where $b_{k}^{\dag }$ ($b_{k}$) is the creation (annihilation)
operator of environmental bosonic mode with frequency $\omega _{k}$,
$\sigma _{x}$ and $\sigma _{z}$ are Pauli matrices to describe the
two-level system. The competing interactions in SBM are between the
quantum tunneling $\Delta$ and the dissipative coupling $g_k$ to the
environment. The effect of the environment is characterized by a
spectral density
$J(\omega)=\sum_kg^2_k\delta(\omega-\omega_k)=2\alpha
\omega^s\omega^{1-s}_c\theta(\omega_c-\omega)$ with the
dimensionless coupling strength $\alpha$ and the hard upper cutoff
at $\omega_c$. The index $s$ accounts for various physical
situations\cite{rmp,book}: the Ohmic $s=1$, sub-Ohmic $s<1$ and
super-Ohmic $s>1$ baths.

The quantum critical point (QCP) and the quantum phase transition
(QPT) are related to the ground state transition, which is usually
triggered by competing interactions. As for SBM, the interesting
phase transition is related to the transition of degeneracy of the
ground state, that is, it is a transition between the non-degenerate
and degenerate ground state\cite{rmp,book,bu1,bu2,hur}. The main
theoretical interest of the QCP in SBM is to understand how the
competing interactions influences the degeneracy of the ground
state. Since the Hamiltonian (1) is invariant under $\sigma_z\to
-\sigma_z$ (together with $b_k, b^{\dag}_k\to -b_k, -b^{\dag}_k$)
and one must have $\langle\sigma_{z}\rangle_G=0$ ($\langle ...
\rangle_G$ means the ground state average). However, for the Ohmic
bath $s=1$ it is well known\cite{rmp,book} that a
Kosterlitz-Thouless quantum transition separates a degenerate ground
state at $\alpha>\alpha_c$ from a non-degenerate one at
$\alpha<\alpha_c$ ($\alpha_c=1$ in the scaling limit $\Delta\ll
\omega_c$).

The ground state of SBM Hamiltonian (1) was studied by many authors
using various analytic and numerical methods. Silbey and Harris
(SH)\cite{sh} proposed an variational ground state and predicted the
QCP $\alpha_c=1$ for $s=1$. The SH ground state was used by Kehrein
and Mielke\cite{km} for sub-Ohmic ($s<1$) bath to calculate the QCP
$\alpha_c$. In last ten years, various numerical techniques were
used for calculation of the QCP in the SBM, such as the numerical
renormalization group (NRG)\cite{bu1,bu2,hur}, the quantum Monte
Carlo (QMC)\cite{qmc}, the method of sparse polynomial space
representation\cite{fehske}, the extended coherent state
approach\cite{chen}, and the variational matrix product state
approach\cite{vmps}. Besides, recently an extension of the
Silbey-Harris ground state was proposed by Zhao {\it et
al.}\cite{zhao} and Chin {\it et al.}\cite{chin} to study the QPT in
the $s=1/2$ sub-Ohmic SBM.

In this work, we propose an analytic ground state wavefunction for
the SBM, which is non-Gaussian for the bath modes and is an
extension of the work of Zhao {\it et al.}\cite{zhao} and Chin {\it
et al.}\cite{chin}. The QPT is usually not a weak coupling problem
and people believe that the numerical techniques may be more
powerful than approximate analytic methods for strong coupling
problem. Then, why do we still try to find an approximate analytical
solution? Generally speaking, our purpose is to see and understand
the physics more clearly and straightforwardly. In particular, here
our purpose is to understand the role played by the infrared
divergence in the SBM Hamiltonian (1).

The QPT in quantum impurity systems may be related to the infrared
catastrophe in baths. P. W. Anderson\cite{and} was the first to
point out this relation for the Anderson model and Kondo model in
fermionic bath. Our question is: What is the role played by the
infrared catastrophe in the quantum phase transition in bosonic bath
of SBM?

\section{The ground state}

If $\Delta=0$, Hamiltonian (1) is solvable and we have degenerate
ground state
\begin{eqnarray}
&&|\psi_{\uparrow(\downarrow)}\rangle=\exp[-\sum_kg_k(b_{k}^{\dag
}-b_{k})\sigma_z/2\omega_k]|\uparrow(\downarrow)\rangle|\{0_k\}\rangle,
\end{eqnarray}
where $|\uparrow(\downarrow)\rangle $ is the eigenstate of
$\sigma_z$: $\sigma_z|\uparrow(\downarrow)\rangle
=+(-)|\uparrow(\downarrow)\rangle$ and $|\{0_{k}\}\rangle$ is the
vacuum state of the bath. Then, for finite $\Delta$ it is naturally
to use a superposed ground state to remove the degeneracy. But it is
well known\cite{rmp,book,sh,km} that there exists an infrared
divergence in the overlap between the degenerate states:
$\langle\psi_{\uparrow}|\psi_{\downarrow}\rangle=\exp[-\sum_kg^2_k/2\omega^2_k]=0$
for $s\le 1$. Silbey and Harris proposed a modified superposed
ground state\cite{sh}
\begin{eqnarray}
&&|G_{SH}\rangle=\exp[-\sum_kg_k(b_{k}^{\dag
}-b_{k})\sigma_z/2(\omega_k+\eta_0\Delta)]2^{-1/2}(|\uparrow\rangle+|\downarrow\rangle)|\{0_k\}\rangle,
\end{eqnarray}
with finite renormalized overlap
$\eta_0=\exp[-\sum_kg^2_k/2(\omega_k+\eta_0\Delta)^2]$ where the
infrared divergence has been removed. The ground state energy is
\begin{eqnarray}
&&E^{SH}_g=-\eta_0\Delta/2-\sum_kg^2_k(\omega_k+2\eta_0\Delta)/4(\omega_k+\eta_0\Delta)^2.
\end{eqnarray}
For the SH ground state at the scaling limit $\Delta\ll \omega_c$,
$\eta_0=(e\Delta/\omega_c)^{\frac{\alpha}{1-\alpha}}$ for $s=1$ and
thus the QCP is at $\alpha^{SH}_c=1$ where $\eta_0=0$. For sub-ohmic
bath $s<1$ one can calculate the QCP by condition: $\eta_0=0$ at
$\alpha\rightarrow\alpha^{SH}_c$\cite{km,lu}, and some results are
listed in the second column of table 1.

Zhao {\it et al.}\cite{zhao} and Chin {\it et al.}\cite{chin}
proposed an extension of the Silbey-Harris ground state to study the
QPT in the $s=1/2$ sub-Ohmic spin-boson model, with degenerate
ground state when zero-biased and $\alpha>\alpha^{D}_c$ (superscript
"D" means degenerate)
\begin{eqnarray}
&&|\Psi_{\pm}\rangle=\exp(-S_{\pm})(u_{\pm}|\uparrow\rangle+v_{\pm}|\downarrow\rangle)|\{0_k\}\rangle,\\
&&S_{\pm}=\sum_{k}\frac{g_{k}}{2\omega _{k}}(b_{k}^{\dag
}-b_{k})[\xi _{k}\sigma _{z}{\pm}(1-\xi _{k})\phi_k],
\end{eqnarray}
where $u_+=v_-=2^{-1/2}\sqrt{1+M}$, $u_-=v_+=2^{-1/2}\sqrt{1-M}$,
$\xi_k=\omega_k/(\omega_k+W)$, $W=\eta \Delta/\sqrt{1-M^2}$, and
\begin{eqnarray}
&&\eta =\exp
\left[-\sum_{k}g_{k}^{2}\xi _{k}^{2}/2\omega _{k}^{2}\right],\\
&&M=\sum_k g_{k}^{2}\phi_k(1-\xi _{k})^{2}/(\omega _{k}W).
\end{eqnarray}
Zhao {\it et al.}\cite{zhao} and Chin {\it et al.}\cite{chin} let
$\phi_k=M$ to be a constant in Eqs.(6) and (8), thus
$|\Psi_+\rangle$ and $|\Psi_-\rangle$ are degenerate with degenerate
ground state energy
\begin{eqnarray}
&&E^{D}_g=-W/2-\sum_kg^2_k\xi_k(2-\xi_k)/4\omega_k+\sum_kg^2_kM^2(1-\xi_k)^2/4\omega_k.
\end{eqnarray}
Besides, the average
$\langle\Psi_{\pm}|\sigma_z|\Psi_{\pm}\rangle=\pm M$ may be finite.
They proposed that the QCP is at $\alpha=\alpha^D_c$ where a nonzero
$M$ leads to lower ground state energy (Note that when
$\alpha\le\alpha^{D}_c$, $M=0$ and
$|\Psi_+\rangle=|\Psi_-\rangle=|G_{SH}\rangle$ ). Some $\alpha^D_c$
values for different baths are listed in the third column of table
1. But, as mentioned above, since the Hamiltonian (1) is invariant
under $\sigma_z\to -\sigma_z$ (together with $b_k, b^{\dag}_k\to
-b_k, -b^{\dag}_k$) we should have $\langle\sigma_{z}\rangle_G=0$.

\section{The infrared catastrophe}

The wavefunction of every bath mode in $|G_{SH}\rangle$ or
$|\Psi_{\pm}\rangle$ is a Gaussian function, thus these ground
states are in the Gaussian approximation. Following the proposal of
Shore and Sander\cite{ss} we propose the following superposed ground
state for the SBM, which is beyond the Gaussian approximation and
takes into account the effect of quantum fluctuations,
\begin{eqnarray}
&&|G\rangle=A(|\Psi_+\rangle+|\Psi_-\rangle),
\end{eqnarray}
where $A$ is a normalization factor. Then, it is easy to check that
$\langle G|\sigma_z|G\rangle=0$. But if one choose $\phi_k=M$ in
Eqs.(6) and (8), as was pointed out by Chin {\it et al.}\cite{chin},
there is an infrared divergence of the occupation number of the
nonadiabatic (NA) modes. We show that this divergence leads to the
orthogonality catastrophe between $|\Psi_+\rangle$ and
$|\Psi_-\rangle$,
\begin{eqnarray}
&&\rho=\langle
\Psi_-|\Psi_+\rangle=\langle\{0_{k}\}|\exp\left(-\sum_k\frac{g_k}{\omega_k}(1-\xi_k)(b^{\dag}_k-b_k)M\right)|\{0_{k}\}\rangle
\nonumber\\
&&=\exp\left(-\sum_k\frac{g^2_k}{2\omega^2_k}(1-\xi_k)^2M^2\right)
=\exp\left(-\alpha M^2
W^2\int_0\frac{\omega^{s-2}d\omega}{(\omega+W)^2}\right)=0
\end{eqnarray}
for Ohmic ($s=1$) and sub-Ohmic ($s<1$) baths as the integration in
the exponential is infrared divergent. This is similar to the
infrared catastrophe in Fermi sea interacting with a quantum
impurity\cite{and}. Because of the orthogonality catastrophe the
ground states, $|G_D\rangle=|\Psi_+\rangle$,
$|G_D\rangle=|\Psi_-\rangle$ or $|G\rangle$ (Eq.(10)) are degenerate
with ground state energy (9).

The way to avoid the infrared catastrophe is similar to the proposal
of Anderson\cite{and}, that is, quantum fluctuation of the NA modes
leads to a $k$-dependent $\phi_k$ in Eqs.(6) and (8) removing the
infrared divergence. Then the ground state energy $E_g$ of the
superposed ground state (10) is
\begin{eqnarray}
&&E_g=(E_0+\rho U)/(1+\rho\sqrt{1-M^2}),
\end{eqnarray}
where $E_0=\langle \Psi_+|H|\Psi_+\rangle$, $\rho=\langle
\Psi_+|\Psi_-\rangle$ and $\rho U=\langle \Psi_+|H|\Psi_-\rangle$.
Here
\begin{eqnarray}
&&E_0=-W/2-\sum_{k}g_{k}^{2}\xi _{k}(2-\xi
_{k})/4\omega _{k}+Y,\\
&&U=\sqrt{1-M^2}\left(-\eta^2\Delta^2/2W-\sum_{k}g_{k}^{2}\xi _{k}(2-\xi_{k})/4\omega _{k}-Y\right)\nonumber\\
&&-\eta\Delta\left[\cosh(M)-1-M(\sinh(M)-M)\right]/2,
\end{eqnarray}
and $Y=\sum_{k}g_{k}^{2}\phi_k^{2}(1-\xi _{k})^{2}/4\omega _{k}$.

The variational function $\phi_k$ can be determined by
\begin{eqnarray}
&&\frac{\partial E_g}{\partial \phi_k}=0=\frac{\partial
E_g}{\partial M}\frac{\partial M}{\partial \phi_k}+\frac{\partial
E_g}{\partial \rho}\frac{\partial \rho}{\partial
\phi_k}+\frac{\partial E_g}{\partial Y}\frac{\partial Y}{\partial
\phi_k}
\end{eqnarray}
for every mode $k$ and the result is
\begin{eqnarray}
&&\phi_k=\tau\omega_k/(\omega_k+\rho\delta),
\end{eqnarray}
where $\delta=2(E_0\sqrt{1-M^2}-U)/[(1-\rho)(1+\rho\sqrt{1-M^2})]$
and $\tau$ is the variational parameter. In this way, the
overlapping integral is
\begin{eqnarray}
&&\rho=\exp\left(-\sum_k\frac{g^2_k}{2\omega^2_k}(1-\xi_k)^2\phi^2_k\right)
=\exp\left(-\alpha \tau^2 W^2\int_0\frac{\omega^s
d\omega}{(\omega+W)^2(\omega+\rho\delta)^2}\right),
\end{eqnarray}
which is finite as long as $s>0$.

For $s=1$ the result of variational calculation is shown in Fig.1.
When $\alpha$ goes to $1$, the variational parameter $\tau$ tends to
$1$ and the overlapping $\rho$ decreases to zero as follows:
\begin{eqnarray}
&&\rho=\left[\frac{\delta}{W}\right]^{\frac{\alpha\tau^2}{1-\alpha\tau^2}}
\exp\left(\frac{\alpha\tau^2}{1-\alpha\tau^2}\left[\ln(1+W/\omega_c)+\frac{2+W/\omega_c}{1+W/\omega_c}\right]\right),
\end{eqnarray}
that is, $\rho\to 0$ when $\alpha\to 1-0^+$ ($0^+$ is a positive
infinitesimal) since $\tau\to 1$. This is to say that for $s=1$ the
ground state becomes doubly degenerate when $\alpha\to \alpha_c=1$.

$\alpha_c$ for sub-Ohmic bath ($s<1$) can be calculated in the
similar way, that is, the QCP where the ground state changes from
non-degenerate ($\alpha<\alpha_c$) to doubly degenerate
($\alpha>\alpha_c$). Our results for some $s$ values are shown in
Table 1. For comparison, the numerical results by NRG\cite{bu1}, by
QMC\cite{qmc}, by the method of sparse polynomial space
representation\cite{fehske}, and by the extended coherent state
approach\cite{chen} are also shown. One can see that our result
compare well with these numerical results.

Fig.2 shows the difference between our calculation of the ground
state energy and that of Zhao {\it et al.}\cite{zhao} and Chin {\it
et al.}\cite{chin}, $\delta E_g=E_g-E^{D}_g$. The lower ground state
energy indicates that the ansatz of this work is a better one for
the real ground state.

We note that when $s>1$ (super-Ohmic bath) the overlapping $\rho$ in
Eq.(17) has always a finite solution. This is to say that the ground
state of the SBM with super-Ohmic bath is always non-degenerate and
there is no QPT.

\section{Conclusion}

We propose an analytic ground state wavefunction for the unbiased
spin-boson Hamiltonian, which is a superposition of the two
degenerate state and is non-Gaussian for the bosonic bath modes. The
infrared catastrophe in Ohmic and sub-Ohmic bosonic bath plays an
important role in determining the degeneracy of the ground state and
we show that the infrared divergence associated with the
displacement of the nonadiabatic modes in bath may be removed from
the proposed ground state for the coupling $\alpha<\alpha_c$. The
QCP $\alpha_c$ is determined by the transition from non-degenerate
to degenerate ground state.  Our ground state energy is lower than
previous authors'results. The calculation of $\alpha_c$ agrees well
with previous numerical results.

\vskip 0.5cm

{\noindent {\large {\bf Acknowledgement}}}

We are grateful to H. Rieger's group and H. Fehske's group for
providing numerical data for comparison. We are also grateful to
Qinghu Chen, A. Winter, F. Anders, A. Alvermann, P. Nalbach,
Ning-Hua Tong, Cheng Guo, Ding Ping Li and Weimin Zhang for
discussions. This work was supported by the National Natural Science
Foundation of China (Grant No. 11174198, 10904091 and 91221201) and
the National Basic Research Program of China (Grant No.
2011CB922202).

\newpage

\vskip 0.5cm

\baselineskip 20pt

\begin{figure}[h]
\centering
  \includegraphics[width=4in]{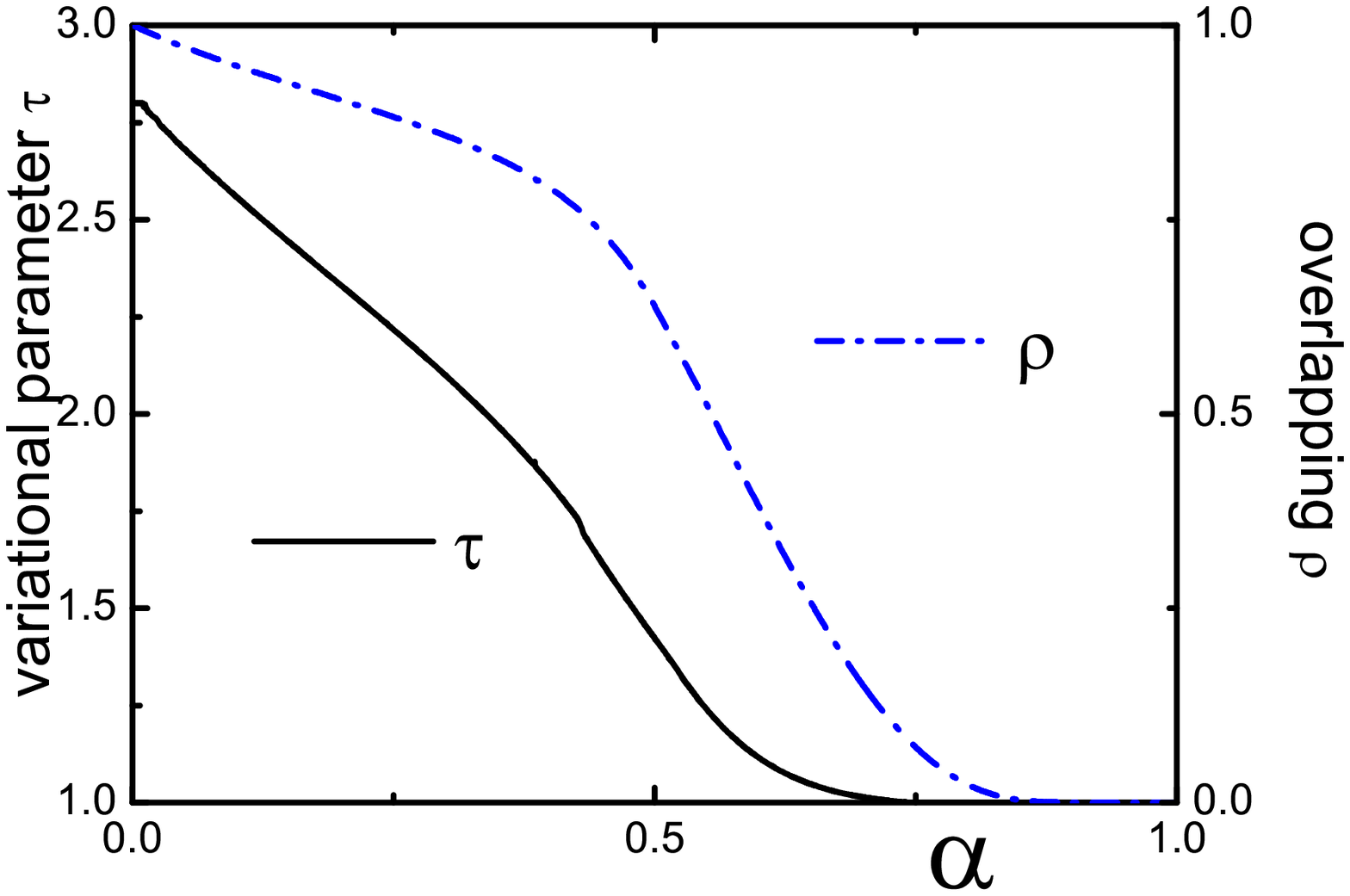}\\
  \caption{The variation parameter $\tau$ and the overlapping
$\rho$ as functions of $\alpha$ for Ohmic bath $s=1$. }
\end{figure}

\newpage

\begin{figure}[h]
\centering
  \includegraphics[width=4in]{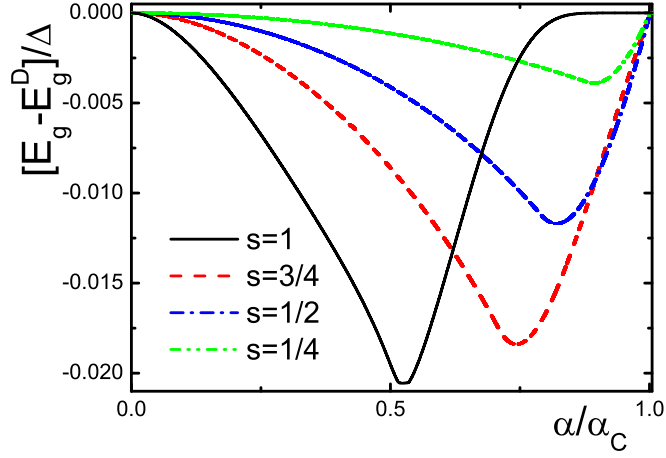}\\
  \caption{$E_g-E^D_g$ is the difference between ground state
energies calculated by Eq.(12) and Eq.(9). $\Delta/\omega_c=0.1$.
See text for details. }
\end{figure}

\vskip 1cm

\begin{center}
{\Large \bf Tables }
\end{center}

{\bf Table 1}~~~QCP of different bath type $s$.

\begin{table}[h]
\centering\setlength\tabcolsep{10pt}
\begin{tabular}{|cccccccc|}
\hline
 $s$ & $\alpha^{SH}_c$ & $\alpha^D_c$ & Our $\alpha_c$
& $\alpha_c$\cite{bu1} & $\alpha_c$\cite{qmc} &
$\alpha_c$\cite{fehske} & $\alpha_c$\cite{chen}
\\ \hline
 1/4    &   0.08554 &  0.02413  &    0.02744  &  0.0264  &  0.0254  &  0.0259 &  0.0256 \\
 1/2    &   0.1768  &  0.08555  &    0.1084   &  0.1065  &  0.0983  &  0.0977 &  0.0820 \\
 3/4    &   0.3537  &  0.2176   &    0.3076   &  0.3168  &  0.2951  &  0.2953 &  0.3205 \\
 1      &   1       &  0.5121   &    1        &   1  &   1   &  1   & 1  \\
\hline

\end{tabular}

\end{table}

\end{document}